\documentclass[conference]{IEEEtran}
\IEEEoverridecommandlockouts
\usepackage{cite}
\usepackage{amsmath,amssymb,amsfonts}
\usepackage{graphicx}
\usepackage{textcomp}
\usepackage{xcolor}
\usepackage{stfloats}
\usepackage{url}
\usepackage{verbatim}
\usepackage{cite}
\usepackage[ruled,vlined,linesnumbered]{algorithm2e}
\usepackage{algpseudocode}
\usepackage{multirow}
\usepackage{amsfonts}
\usepackage{xspace}
\usepackage{colortbl}
\usepackage{makecell}
\usepackage{microtype}
\usepackage{enumitem}
\usepackage{pifont}
\usepackage{booktabs}
\usepackage{listings}
\usepackage{hyperref}
\usepackage[symbol]{footmisc}

\newcommand{\tool}{\textit{DriveTester}\xspace}
\newcommand{\runner}{\textit{Scenario Runner}\xspace}
\newcommand{\testing}{\textit{Testing Engine}\xspace}

\def\BibTeX{{\rm B\kern-.05em{\sc i\kern-.025em b}\kern-.08em
    T\kern-.1667em\lower.7ex\hbox{E}\kern-.125emX}}
\begin{document}

\title{\tool: A Unified Platform for Simulation-Based Autonomous Driving Testing}

\author{Mingfei Cheng \\
\IEEEauthorblockA{
\textit{Singapore Management University}\\
Singapore
}

\and
\IEEEauthorblockN{Yuan Zhou}
\IEEEauthorblockA{
\textit{Zhejiang Sci-Tech University}\\
China}
\and
\IEEEauthorblockN{Xiaofei Xie}
\IEEEauthorblockA{
\textit{Singapore Management University}\\
Singapore}
}

\maketitle

\begin{abstract}
Simulation-based testing plays a critical role in evaluating the safety and reliability of autonomous driving systems (ADSs). However, one of the key challenges in ADS testing is the complexity of preparing and configuring simulation environments, particularly in terms of compatibility and stability between the simulator and the ADS. This complexity often results in researchers dedicating significant effort to customize their own environments, leading to disparities in development platforms and underlying systems. Consequently, reproducing and comparing these methodologies on a unified ADS testing platform becomes difficult.
To address these challenges, we introduce \tool, a unified simulation-based testing platform built on Apollo, one of the most widely used open-source, industrial-level ADS platforms. \tool provides a consistent and reliable environment, integrates a lightweight traffic simulator, and incorporates various state-of-the-art ADS testing techniques. This enables researchers to efficiently develop, test, and compare their methods within a standardized platform, fostering reproducibility and comparison across different ADS testing approaches.
The code is available: \href{https://github.com/MingfeiCheng/DriveTester}{https://github.com/MingfeiCheng/DriveTester}.

\end{abstract}


\section{Introduction}
\footnotetext{Ongoing work (v1): initial version}
Autonomous Driving Systems (ADS) are a groundbreaking technology that enables vehicles to operate without human intervention. They rely on a combination of sensors 
and artificial intelligence algorithms to perceive the environment, make decisions, and navigate safely. Ensuring the safety and reliability of ADS before deployment is critical, making thorough testing an essential part of their development.

A significant body of work has been conducted on ADS testing, with many approaches relying on \textit{simulation} due to the high costs and risks associated with real-world vehicle testing. However, one of the key challenges in simulation-based ADS testing is the preparation and configuration of simulation environments, which is often time-consuming, complex, and highly dependent on specific platforms. The frequent updates to simulators, ADS software, and testing methods make it difficult to maintain a reliable testing platform, limiting existing ADS testing tools to be easily reproduced and compared on a unified platform. For instance, Baidu Apollo~\cite{apollo}, one of the most widely studied open-source industrial-level ADS platforms, has traditionally been tested using the LGSVL simulator~\cite{rong2020lgsvl}. However, since LGSVL was sunsetted in 2022~\cite{LGSVLSunsetting}, compatibility issues between Apollo and alternative simulators, such as CARLA~\cite{dosovitskiy2017carla}, have posed significant challenges for researchers, complicating the replication, development, and comparison of various ADS testing techniques. Furthermore, many existing ADS testing tools are built with simulator-specific APIs, making it difficult to transfer these tools across different platforms. For example, migrating SAMOTA, a tool developed based on the Pylot platform~\cite{gog2021pylot}, to Apollo requires substantial manual effort.

\begin{figure}
    \centering
    \includegraphics[width=1.0\linewidth]{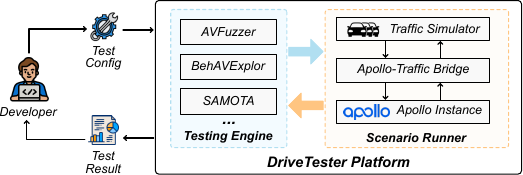}
    \vspace{-20pt}
    \caption{A high-level overview of \textit{\tool}.}
    \label{fig:overview}
    \vspace{-20pt}
\end{figure}

In this paper, to address the aforementioned limitations, we introduce \tool, a simulation-based testing platform for Apollo. \tool provides a unified, extensible, and user-friendly environment for ADS testing, integrating multiple state-of-the-art ADS testing baseline methods. This platform aims to simplify the testing process, enhance reproducibility, and facilitate direct comparisons across different ADS testing techniques.
As shown in Fig.~\ref{fig:overview}, \tool consists of two main components: \testing and \runner. The \testing module integrates state-of-the-art ADS testing techniques (e.g., AVFuzzer~\cite{av_fuzzer}), which automatically identify violations (e.g., collisions) for Apollo. Note that migrating these techniques to \tool involved preserving their core algorithmic components, a process that required considerable development effort. Researchers can easily select their desired testing technique using a simple configuration interface.
The \runner module is responsible for simulating the specific scenarios generated by \testing. To address compatibility issues with Apollo, we developed a lightweight Traffic Simulator and an Apollo-Traffic Bridge. The Traffic Simulator dynamically generates real-time driving observations (i.e., the states of non-player character (NPC) vehicles) based on the scenario configurations. These observations are continuously updated through interactions with Apollo, facilitated by the Apollo-Traffic communication bridge.

By offering a unified and extensible platform, \tool significantly reduces the complexity and time required for setting up and configuring simulation environments. This allows researchers, particularly from the software engineering community, to focus more on methods evluation and innovation, rather than dealing with the intricacies of environment configuration.


We evaluated the practical utility of \tool by applying it to various existing ADS testing approaches. Our experiments demonstrated that \tool introduces minimal latency from the simulator while effectively detecting critical violations, such as collisions. Furthermore, \tool supports large-scale simulation experiments through parallel and scalable execution of test cases, enabling more efficient resource utilization and faster testing. This capability is particularly advantageous for search-based approaches, which benefit from running multiple test cases with simple configurations.

\section{Related Work}
\noindent \textbf{Autonomous Driving Systems.} 
Existing Autonomous Driving Systems (ADSs), including End-to-End (E2E) systems \cite{openpilot} and module-based ADSs \cite{apollo}, have shown effectiveness in recent studies. In comparison to E2E systems, module-based ADSs, such as Apollo \cite{apollo}, consist of complex components like localization, perception, prediction, planning, and control, along with sophisticated communication mechanisms. While these components enhance real-time performance, they also complicate deployment.
In this paper, we focus on Apollo, an open-source, industrial-level ADS. More importantly, it faces significant challenges in compatibility and usability during ADS testing, which must be addressed by the software engineering research community.


\noindent \textbf{Simulation-based ADS Testing.} Many simulation-based testing approaches have been proposed to evaluate the performance (i.e., safety) of ADSs. These approaches mainly design various algorithms~\cite{huai2023sceno, css_drivefuzzer, huai2023doppelganger, haq2023many, cheng2023behavexplor, cheng2024evaluating, hildebrandt2023physcov,gambi2019automatically,han2021preliminary,av_fuzzer,icse_samota,tse_adfuzz,tang2021systematic,zhou2023specification,tang2021route,tang2021collision,li2023generative,zhang2023building,deng2022scenario,gambi2019generating,najm2013depiction,nitsche2017pre,roesener2016scenario, paardekooper2019automatic,lu2024diavio}. such as guided fuzzing, to explore the scenario space and identify scenarios that violate testing requirements (e.g., reach destination safely). 
However, one of the biggest challenges for researchers is the lack of a unified platform for comparing different algorithms across existing methods. For instance, AVFuzzer was developed on LGSVL+Apollo, while SAMOTA was developed on CARLA+Pylot, making direct comparisons difficult. 
In \textit{\tool}, we invested significant effort to migrate existing ADS testing algorithms into a unified platform, simplifying the process for researchers to use and facilitating their onboarding into the ADS testing domain.



\noindent \textbf{Simulation Environment.} 
The simulator is a crucial component of simulation-based ADS testing, as it provides the environment to deploy ADSs on virtual vehicles. The dominant simulators used in existing ADS testing studies are LGSVL~\cite{rong2020lgsvl} and CARLA~\cite{dosovitskiy2017carla}. However, CARLA does not yet provide a stable bridge for Apollo~\cite{apollo_carla_issue}, and the sunset of LGSVL~\cite{LGSVLSunsetting} presents challenges for continuing to use existing testing techniques on Apollo. Although recent studies~\cite{huai2023sceno, huai2023doppelganger, cheng2024evaluating} have employed simControl~\cite{apollo} as an alternative, its lack of a dynamic vehicle model makes the testing system incomplete, and it sometimes suffers from unstable motion issues. 
Furthermore, most existing ADS testing techniques overlook the perception module due to the latency between simulators and the ADS, demonstrating a low requirement for real-time 3D rendering. To address this, we designed a lightweight Traffic simulator that incorporates virtual dynamic models and generates driving observations without the need for physical rendering. This approach is highly efficient for testing complete ADS decision-making modules, including routing, prediction, planning, and control.

\section{Architecture of \tool}

\tool\ is designed to be modular, extensible, and to automate all key steps in ADS testing, allowing researchers to easily run different testing algorithms or develop new ones to generate desired scenarios. As shown in Fig.~\ref{fig:overview}, users need to provide a simple \textit{Test Configuration} file, then \tool\ will deliver the testing results. Specifically, the \textit{Test Configuration} defines the \textit{Testing Algorithm} and its parameters. The \testing module configures the selected testing tool based on this input file and iteratively generates \textit{Scenario Configurations} for evaluating the ADS. The \runner module then parses the \textit{Scenario Configuration} and prepares the simulation environment for scenario execution. Once the environment is set up, with all necessary NPC actors and Apollo instances in place, the \textit{Traffic simulator} dynamically updates the states (e.g., location, speed, and acceleration) of all participants according to the behaviors defined in the scenario.
During execution, the \textit{Test Recorder} monitors Apollo’s behavior against predefined oracles and logs all traffic data for future analysis. Each module will be explained in the subsequent subsections.

\subsection{Test Configuration}
The core idea of ADS testing is to automatically generate scenarios by designing various algorithms within a defined scenario space. To achieve this, \textit{\tool} models ADS testing with specific configurations and employs a YAML structure to describe the test.

\subsubsection{System Setting}  
This setting defines the parameters and configuration files required for \textit{\tool} to initialize, including the root paths for both \textit{\tool} and Apollo.

\subsubsection{Scenario Setting}  
This setting defines the scenario space for running ADS tests, including details such as the map name and the initial scenario configuration (e.g., the start position and the destination of the ADS under test).

\subsubsection{Scenario Runner Setting} This section focuses on configuring the simulator parameters, such as the simulator platform and saving simulation recordings or not. We designed this part to enable the future integration of various scenario runners across different simulators, such as CARLA.

\subsubsection{Testing Engine Setting}  
This section specifies the testing algorithm to be used (e.g., AVFuzzer) along with the necessary parameters for the selected algorithm. Users can also configure the testing oracles in this section (e.g., collision oracle).

\begin{figure}[!t]
    \centering
    \includegraphics[width=0.6\linewidth]{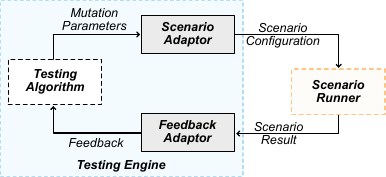}
    \vspace{-10pt}
    \caption{Illustration of \testing}
    \label{fig:te}
    \vspace{-10pt}
\end{figure}
\subsection{\testing}
One of the most challenging aspects of configuring and utilizing existing ADS testing techniques is managing compatibility issues. To overcome this, as shown in Fig.~\ref{fig:te}, we developed two adaptors: the \textit{Scenario Adaptor} and the \textit{Feedback Adaptor}. They enable flexible migration of existing ADS testing techniques from other platforms to \tool\ without altering their core algorithms. Additionally, they provide a flexible foundation for the development of new testing algorithms for further innovation and experimentation. 


\subsubsection{Scenario Adaptor} Since testing algorithms require various mutation parameters to generate new scenarios, we designed the \textit{Scenario Adaptor} to standardize and align the mutation parameters with the features supported by the simulator. For example, the adaptor must parse mutation operators generated by the algorithm (e.g., adding a bicycle in front of the ego vehicle) and map these operations to specific configurations supported by the target simulator (e.g., Traffic Simulator).

\subsubsection{Feedback Adaptor} Similarly, this adaptor is designed to analyze execution recordings from the simulator and produce the feedback format (e.g., ego trajectory) required by the testing algorithms to generate the next set of scenarios.

To reduce the effort required by researchers in developing adaptors for existing ADS testing techniques, \textit{\tool} integrates Random and four state-of-the-art tools, including AVFuzzer~\cite{av_fuzzer}, BehAVExplor~\cite{cheng2023behavexplor}, SAMOTA~\cite{icse_samota}, and DriveFuzz~\cite{css_drivefuzzer}. We are working to include more tools in the future.

\subsection{\runner}
\begin{figure}[!t]
    \centering
    \includegraphics[width=0.9\linewidth]{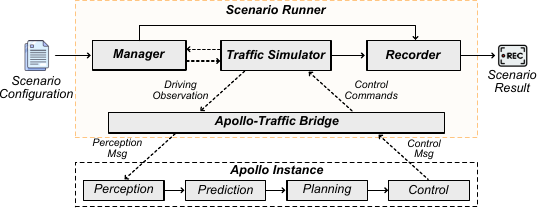}
    \vspace{-10pt}
    \caption{Workflow of \runner}
    \label{fig:runner}
    \vspace{-20pt}
\end{figure}
Fig.~\ref{fig:runner} shows the workflow of \textit{\runner} to execute a scenario, which takes the \textit{Scenario Configuration} as input and outputs a \textit{Scenario Recording}. 

\subsubsection{Scenario Configuration} This configuration defines the positions and behaviors of traffic participants, including vehicles, pedestrians, and static obstacles, within the simulation. For instance, vehicles are expected to follow predefined waypoints as specified in the configuration. The configuration is stored in JSON format for improved reusability and flexibility.

\subsubsection{Manager} This module parses the scenario configuration and creates all necessary actors within the simulator. For autonomous vehicles (e.g., actors controlled by Apollo), it equips a real-time monitor that includes testing oracles to track the state of the vehicles. The simulation automatically terminates when an oracle is violated, enabling efficient time management during testing.

\subsubsection{Recorder} This module automatically starts when the simulation begins and is designed to record the states of all traffic actors at each timestamp. The recorded data is then available for offline analysis, allowing researchers to perform in-depth post-simulation evaluations.

\subsubsection{Traffic Simulator} We designed a real-time simulator that generates traffic behavior satisfying motion constraints. At each timestamp, an actor receives a control command from its policy (e.g., Apollo and PID controllers) and updates its state (i.e., location, heading, speed, and acceleration) based on a rule-based dynamic model, e.g., kinematic equations. 


\subsubsection{Apollo-Traffic Bridge} This module establishes a real-time communication bridge between the simulator and Apollo. The bridge converts traffic observations into perception messages, which are transmitted to Apollo's perception message channel. It listens for control messages from Apollo, extracts the control commands, and transmits them to the simulator. These commands are then used to update the states of the actors controlled by Apollo within the simulation.



\textbf{Discussion.} \textit{\tool} currently provides a complete testing pipeline for Apollo. Our platform is extendable to support new testing algorithms, integrate with different simulators by developing corresponding scenario runners, and accommodate additional ADSs by building the necessary communication bridges. We will continue to maintain the \textit{\tool} repository and plan to support more simulators, ADSs, and testing algorithms.

\section{Usage of \tool}
\textbf{Command Line Interface}. Currently, \textit{\tool} is built for Baidu Apollo, which requires Apollo to be installed first. Additionally, \textit{\tool} is developed entirely in Python, so proper setup and configuration of the Python environment are required. We provide quick installation steps as outlined below:
\begin{lstlisting}[language=bash, frame=single]
bash scripts/setup_DriveTester.sh
\end{lstlisting}

Once the environment and dependencies are set up, users can test Apollo by selecting and configuring the \textit{Test Configuration}. A sample YAML file is provided below to demonstrate how to test Apollo using AVFuzzer:
\begin{lstlisting}[language=bash, frame=single]
# System Setting
system:
  debug: true
  resume: true
# Scenario Setting
scenario:
  map_name: borregas_ave
  start_lane_id: lane_31
  end_lane_id: lane_15
# Scenario Runner Setting
scenario_runner:
  name: ApolloSim
  parameters:
    container_name: null
    save_traffic_recording: true
# Testing Engine Setting
testing_engine:
  # Algorithm Setting
  algorithm:
    name: avfuzzer
    parameters:
      run_hour: 2
      local_run_hour: 0.5
      population_size: 4
      pm: 0.6
      pc: 0.6
  # Oracle Setting
  oracle:
    collision:
      threshold: 0.01
\end{lstlisting}

\begin{figure}[!h]
    \centering
    \includegraphics[width=1.0\linewidth]{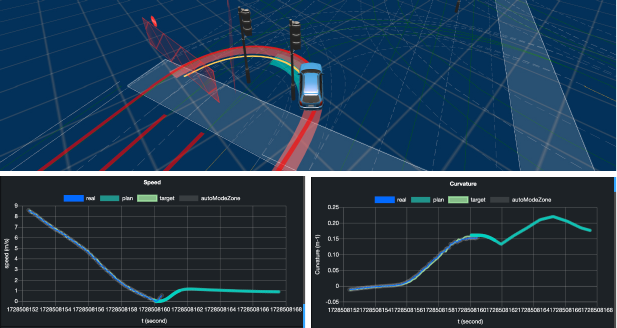}
    \vspace{-20pt}
    \caption{Illustration of correct motion simulation in \textit{\tool}}
    \label{fig:case}
    \vspace{-10pt}
\end{figure}

By using such a \textit{Test Configuration}, users can easily run the ADS test with the following command:
\begin{lstlisting}[language=bash, frame=single]
python main.py -cn avfuzzer
\end{lstlisting}


\textbf{Web-UI Visualization}.  
As shown in Fig.~\ref{fig:case}, Apollo provides Dreamview as a visualization tool, which is also supported by \tool for simulation visualization. 


\begin{table}[!h]
    \centering
    \caption{The effectiveness of \textit{\tool}}
    \vspace{-5pt}
    \small
    \resizebox{\linewidth}{!}{
        \begin{tabular}{l|ccccc}
        \toprule
         {\textbf{Scenarios}} & \textit{Random} & \textit{AVFuzzer} & \textit{BehAVExplor} & \textit{SAMOTA} & \textit{DriveFuzz} \\
         \midrule
        \textit{Turning Left} & 44 & 48 & 55 & 46 & 32\\
        \textit{Crossing Junction} & 40 & 54 & 69 & 50 & 30\\
        \bottomrule
        \end{tabular}
    }
    \label{tab:res}
\end{table}

\section{Evaluation}
We evaluate \textit{\tool} in two ways:
(1) \textbf{Usability.} We aimed to ensure that \textit{\tool} can be easily used on a non-GUI Linux server. To evaluate this, we first confirmed that the system could be installed and run successfully on a non-GUI Linux server. Specifically, we tested it on a Linux server with an AMD EPYC 7543P 32-Core Processor (256GB RAM) and an NVIDIA A5000 GPU. Additionally, we verified that the simulator could accurately simulate vehicle motion states, particularly the motion of Apollo. Fig.~\ref{fig:case} demonstrates a U-turn scenario used to test the dynamic model in \textit{\tool}. As the ego vehicle approached the intersection, the control module applied the brakes and turned the steering wheel left to execute the U-turn. The planned speed and curvature (green lines), along with the simulated vehicle states (blue lines), are displayed in the lower sections of Fig.~\ref{fig:case}. 
(2) \textbf{Effectiveness.} We aimed to ensure that the migrated ADS testing tools can still effectively identify violations when using \textit{\tool}. To verify this, we simply ran each tool for 2 hours on two widely used scenarios, \textit{Turning Lleft} and \textit{Crossing Junction}, and analyzed the results. From the results in TABLE~\ref{tab:res}, we observe that using \textit{\tool} maintains the performance of existing testing tools in identifying violations.



\section{Conclusion}  
In this paper, we introduce \tool, a unified simulation-based platform for ADS testing, aiming to address the challenges of compatibility and environment configuration. Built for the Apollo platform, \tool integrates state-of-the-art ADS testing techniques and simplifies the process of developing and executing tests. By offering modular and extensible components, \tool facilitates seamless migration of existing testing techniques and supports the development of new ones. Our evaluation demonstrates the high usability of \tool and its effectiveness in discovering critical violations.



\bibliographystyle{IEEEtran}
\bibliography{reference}

\begin{thebibliography}{10}
\providecommand{\url}[1]{#1}
\csname url@samestyle\endcsname
\providecommand{\newblock}{\relax}
\providecommand{\bibinfo}[2]{#2}
\providecommand{\BIBentrySTDinterwordspacing}{\spaceskip=0pt\relax}
\providecommand{\BIBentryALTinterwordstretchfactor}{4}
\providecommand{\BIBentryALTinterwordspacing}{\spaceskip=\fontdimen2\font plus
\BIBentryALTinterwordstretchfactor\fontdimen3\font minus \fontdimen4\font\relax}
\providecommand{\BIBforeignlanguage}[2]{{%
\expandafter\ifx\csname l@#1\endcsname\relax
\typeout{** WARNING: IEEEtran.bst: No hyphenation pattern has been}%
\typeout{** loaded for the language `#1'. Using the pattern for}%
\typeout{** the default language instead.}%
\else
\language=\csname l@#1\endcsname
\fi
#2}}
\providecommand{\BIBdecl}{\relax}
\BIBdecl

\bibitem{apollo}
\BIBentryALTinterwordspacing
Baidu, ``Apollo: Open source autonomous driving,'' 2019. [Online]. Available: \url{https://github.com/ApolloAuto/apollo}
\BIBentrySTDinterwordspacing

\bibitem{rong2020lgsvl}
G.~Rong, B.~H. Shin, H.~Tabatabaee, Q.~Lu, S.~Lemke, M.~Mo{\v{z}}eiko, E.~Boise, G.~Uhm, M.~Gerow, S.~Mehta \emph{et~al.}, ``{LGSVL} simulator: A high fidelity simulator for autonomous driving,'' in \emph{2020 IEEE 23rd International Conference on Intelligent Transportation Systems (ITSC)}.\hskip 1em plus 0.5em minus 0.4em\relax Rhodes, Greece: IEEE, 2020, pp. 1--6.

\bibitem{LGSVLSunsetting}
lgsvl, ``{LGSVL Sunsetting},'' \url{https://github.com/lgsvl/simulator}, 2022.

\bibitem{dosovitskiy2017carla}
A.~Dosovitskiy, G.~Ros, F.~Codevilla, A.~Lopez, and V.~Koltun, ``{CARLA}: {An} open urban driving simulator,'' in \emph{Proceedings of the 1st Annual Conference on Robot Learning}, ser. Proceedings of Machine Learning Research, S.~Levine, V.~Vanhoucke, and K.~Goldberg, Eds., vol.~78.\hskip 1em plus 0.5em minus 0.4em\relax California, USA: PMLR, 2017, pp. 1--16.

\bibitem{gog2021pylot}
I.~Gog, S.~Kalra, P.~Schafhalter, M.~A. Wright, J.~E. Gonzalez, and I.~Stoica, ``Pylot: A modular platform for exploring latency-accuracy tradeoffs in autonomous vehicles,'' in \emph{2021 IEEE International Conference on Robotics and Automation (ICRA)}.\hskip 1em plus 0.5em minus 0.4em\relax IEEE, 2021, pp. 8806--8813.

\bibitem{av_fuzzer}
G.~Li, Y.~Li, S.~Jha, T.~Tsai, M.~Sullivan, S.~K.~S. Hari, Z.~Kalbarczyk, and R.~Iyer, ``{AV-FUZZER}: Finding safety violations in autonomous driving systems,'' in \emph{2020 IEEE 31st International Symposium on Software Reliability Engineering (ISSRE)}.\hskip 1em plus 0.5em minus 0.4em\relax Coimbra, Portugal: IEEE, 2020, pp. 25--36.

\bibitem{openpilot}
\BIBentryALTinterwordspacing
comma.ai, ``{OpenPilot}: An open source driver assistance system,'' 2022. [Online]. Available: \url{https://github.com/commaai/openpilot}
\BIBentrySTDinterwordspacing

\bibitem{huai2023sceno}
Y.~Huai, S.~Almanee, Y.~Chen, X.~Wu, Q.~A. Chen, and J.~Garcia, ``sceno rita: Generating diverse, fully-mutable, test scenarios for autonomous vehicle planning,'' \emph{IEEE Transactions on Software Engineering}, 2023.

\bibitem{css_drivefuzzer}
S.~Kim, M.~Liu, J.~J. Rhee, Y.~Jeon, Y.~Kwon, and C.~H. Kim, ``Drivefuzz: Discovering autonomous driving bugs through driving quality-guided fuzzing,'' in \emph{Proceedings of the 2022 ACM SIGSAC Conference on Computer and Communications Security}.\hskip 1em plus 0.5em minus 0.4em\relax Los Angeles, CA, USA: ACM, 2022, pp. 1753--1767.

\bibitem{huai2023doppelganger}
Y.~Huai, Y.~Chen, S.~Almanee, T.~Ngo, X.~Liao, Z.~Wan, Q.~A. Chen, and J.~Garcia, ``Doppelg{\"a}nger test generation for revealing bugs in autonomous driving software,'' in \emph{2023 IEEE/ACM 45th International Conference on Software Engineering (ICSE)}.\hskip 1em plus 0.5em minus 0.4em\relax IEEE, 2023, pp. 2591--2603.

\bibitem{haq2023many}
F.~U. Haq, D.~Shin, and L.~C. Briand, ``Many-objective reinforcement learning for online testing of dnn-enabled systems,'' in \emph{2023 IEEE/ACM 45th International Conference on Software Engineering (ICSE)}.\hskip 1em plus 0.5em minus 0.4em\relax IEEE, 2023, pp. 1814--1826.

\bibitem{cheng2023behavexplor}
M.~Cheng, Y.~Zhou, and X.~Xie, ``Behavexplor: Behavior diversity guided testing for autonomous driving systems,'' in \emph{Proceedings of the 32nd ACM SIGSOFT International Symposium on Software Testing and Analysis}, 2023, pp. 488--500.

\bibitem{cheng2024evaluating}
M.~Cheng, Y.~Zhou, X.~Xie, J.~Wang, G.~Meng, and K.~Yang, ``Evaluating decision optimality of autonomous driving via metamorphic testing,'' \emph{arXiv preprint arXiv:2402.18393}, 2024.

\bibitem{hildebrandt2023physcov}
C.~Hildebrandt, M.~von Stein, and S.~Elbaum, ``Physcov: Physical test coverage for autonomous vehicles,'' in \emph{Proceedings of the 32nd ACM SIGSOFT International Symposium on Software Testing and Analysis}, 2023, pp. 449--461.

\bibitem{gambi2019automatically}
A.~Gambi, M.~Mueller, and G.~Fraser, ``Automatically testing self-driving cars with search-based procedural content generation,'' in \emph{Proceedings of the 28th ACM SIGSOFT International Symposium on Software Testing and Analysis}.\hskip 1em plus 0.5em minus 0.4em\relax Beijing, China: ACM, 2019, pp. 318--328.

\bibitem{han2021preliminary}
S.~Han, J.~Kim, G.~Kim, J.~Cho, J.~Kim, and S.~Yoo, ``Preliminary evaluation of path-aware crossover operators for search-based test data generation for autonomous driving,'' in \emph{2021 IEEE/ACM 14th International Workshop on Search-Based Software Testing (SBST)}.\hskip 1em plus 0.5em minus 0.4em\relax Madrid, Spain: IEEE, 2021, pp. 44--47.

\bibitem{icse_samota}
F.~U. Haq, D.~Shin, and L.~Briand, ``Efficient online testing for dnn-enabled systems using surrogate-assisted and many-objective optimization,'' in \emph{Proceedings of the 44th International Conference on Software Engineering}.\hskip 1em plus 0.5em minus 0.4em\relax Pittsburgh Pennsylvania: IEEE, 2022, pp. 811--822.

\bibitem{tse_adfuzz}
Z.~Zhong, G.~Kaiser, and B.~Ray, ``Neural network guided evolutionary fuzzing for finding traffic violations of autonomous vehicles,'' \emph{IEEE Transactions on Software Engineering}, vol.~49, no.~4, pp. 1860--1875, 2023.

\bibitem{tang2021systematic}
Y.~Tang, Y.~Zhou, T.~Zhang, F.~Wu, Y.~Liu, and G.~Wang, ``Systematic testing of autonomous driving systems using map topology-based scenario classification,'' in \emph{Proceedings of the 36th IEEE/ACM International Conference on Automated Software Engineering (ASE)}.\hskip 1em plus 0.5em minus 0.4em\relax Melbourne, Australia: IEEE, 2021, pp. 1342--1346.

\bibitem{zhou2023specification}
Y.~Zhou, Y.~Sun, Y.~Tang, Y.~Chen, J.~Sun, C.~M. Poskitt, Y.~Liu, and Z.~Yang, ``Specification-based autonomous driving system testing,'' \emph{IEEE Transactions on Software Engineering}, pp. 1--19, 2023.

\bibitem{tang2021route}
Y.~Tang, Y.~Zhou, F.~Wu, Y.~Liu, J.~Sun, W.~Huang, and G.~Wang, ``Route coverage testing for autonomous vehicles via map modeling,'' in \emph{2021 IEEE International Conference on Robotics and Automation (ICRA)}.\hskip 1em plus 0.5em minus 0.4em\relax Xi'an, China: IEEE, 2021, pp. 11\,450--11\,456.

\bibitem{tang2021collision}
Y.~Tang, Y.~Zhou, Y.~Liu, J.~Sun, and G.~Wang, ``Collision avoidance testing for autonomous driving systems on complete maps,'' in \emph{2021 IEEE Intelligent Vehicles Symposium (IV)}.\hskip 1em plus 0.5em minus 0.4em\relax Nagoya, Japan: IEEE, 2021, pp. 179--185.

\bibitem{li2023generative}
Z.~Li, X.~Wu, D.~Zhu, M.~Cheng, S.~Chen, F.~Zhang, X.~Xie, L.~Ma, and J.~Zhao, ``Generative model-based testing on decision-making policies,'' in \emph{2023 38th IEEE/ACM International Conference on Automated Software Engineering (ASE)}.\hskip 1em plus 0.5em minus 0.4em\relax IEEE, 2023, pp. 243--254.

\bibitem{zhang2023building}
X.~Zhang and Y.~Cai, ``Building critical testing scenarios for autonomous driving from real accidents,'' in \emph{Proceedings of the 32nd ACM SIGSOFT International Symposium on Software Testing and Analysis}, 2023, pp. 462--474.

\bibitem{deng2022scenario}
Y.~Deng, X.~Zheng, M.~Zhang, G.~Lou, and T.~Zhang, ``Scenario-based test reduction and prioritization for multi-module autonomous driving systems,'' in \emph{Proceedings of the 30th ACM Joint European Software Engineering Conference and Symposium on the Foundations of Software Engineering}, 2022, pp. 82--93.

\bibitem{gambi2019generating}
A.~Gambi, T.~Huynh, and G.~Fraser, ``Generating effective test cases for self-driving cars from police reports,'' in \emph{Proceedings of the 2019 27th ACM Joint Meeting on European Software Engineering Conference and Symposium on the Foundations of Software Engineering}.\hskip 1em plus 0.5em minus 0.4em\relax Tallinn Estonia: ACM, 2019, pp. 257--267.

\bibitem{najm2013depiction}
W.~G. Najm, S.~Toma, J.~Brewer \emph{et~al.}, ``Depiction of priority light-vehicle pre-crash scenarios for safety applications based on vehicle-to-vehicle communications,'' National Highway Traffic Safety Administration, U.S. Department of Transportation, Washington, DC, Tech. Rep. DOT HS 811 732, Apr. 2013.

\bibitem{nitsche2017pre}
P.~Nitsche, P.~Thomas, R.~Stuetz, and R.~Welsh, ``Pre-crash scenarios at road junctions: A clustering method for car crash data,'' \emph{Accident Analysis \& Prevention}, vol. 107, pp. 137--151, 2017.

\bibitem{roesener2016scenario}
C.~Roesener, F.~Fahrenkrog, A.~Uhlig, and L.~Eckstein, ``A scenario-based assessment approach for automated driving by using time series classification of human-driving behaviour,'' in \emph{2016 IEEE 19th international conference on intelligent transportation systems (ITSC)}.\hskip 1em plus 0.5em minus 0.4em\relax Rio de Janeiro, Brazil: IEEE, 2016, pp. 1360--1365.

\bibitem{paardekooper2019automatic}
J.-P. Paardekooper, S.~Montfort, J.~Manders, J.~Goos, E.~d. Gelder, O.~Camp, O.~Bracquemond, and G.~Thiolon, ``Automatic identification of critical scenarios in a public dataset of 6000 km of public-road driving,'' in \emph{26th International Technical Conference on the Enhanced Safety of Vehicles (ESV)}.\hskip 1em plus 0.5em minus 0.4em\relax Eindhoven, Netherlands: Mira Smart, 2019.

\bibitem{lu2024diavio}
Y.~Lu, Y.~Tian, Y.~Bi, B.~Chen, and X.~Peng, ``Diavio: Llm-empowered diagnosis of safety violations in ads simulation testing,'' in \emph{Proceedings of the 33rd ACM SIGSOFT International Symposium on Software Testing and Analysis}, 2024, pp. 376--388.

\bibitem{apollo_carla_issue}
guardstrikelab, ``{carla apollo bridge issues},'' \url{https://github.com/guardstrikelab/carla_apollo_bridge/issues}, 2024.

\end{thebibliography}

\end{document}